\documentclass[aps,prb,superscriptaddress,twocolumn,tightenlines,showpacs]{revtex4}

\usepackage {graphicx}
\usepackage {amsmath}
\usepackage {amsfonts}
\usepackage {amssymb}
\usepackage {mathrsfs}

\usepackage{epstopdf}

\newcommand {\be}{\begin{eqnarray}}
\newcommand {\ee}{\end{eqnarray}}
\newcommand {\rmd} {{\rm d}}
\newcommand {\pathD} {{\mathscr D}}

\begin{document}

\title {Vortex duality: watching the dual side with order propagators.}
\author {V. Cvetkovic}
\email {vladimir@lorentz.leidenuniv.nl}
\affiliation {Instituut Lorentz voor de theoretische natuurkunde, 
Universiteit Leiden, P.O. Box 9506, NL-2300 RA Leiden, 
The Netherlands}
\author {J. Zaanen}
\email {jan@lorentz.leidenuniv.nl}
\affiliation {Instituut Lorentz voor de theoretische natuurkunde, 
Universiteit Leiden, P.O. Box 9506, NL-2300 RA Leiden, 
The Netherlands}

\date {\today}
\begin {abstract}
In condensed matter physics, Kramers-Wannier duality implies that the
state disordered by quantum fluctuations or temperature actually
corresponds with an ordered state formed from the topological excitations
of the 'original' ordered state. At first sight it might appear to be
impossible to observe this dual order using means associated with the
original order. Although true for Ising models, we consider in this
paper the well known vortex duality, in particular in the quantum
interpretation in 2+1D where it is associated with the quantum phase
transition from a superfluid to a Bose Mott insulator. Here, the
disordered Mott insulating state is at the same time a dual 
superconductor corresponding with a Bose condensate of vortices.
We present a simple formalism making it possible to compute 
the velocity propagator associated with the superfluid in terms
of the degrees of freedom of the dual theory. The Mott insulator
is characterized by a doublet of massive modes and we demonstrate 
that one of these modes is  nothing else than the longitudinal photon
(gauged second sound) of the dual superconductor. For increasing momenta,
the system rediscovers the original order, and the effect on the
velocity correlator is that the longitudinal photon looses its 
pole strength. The quantum critical regime as probed by the velocity
correlator is most interesting. We demonstrate that at infinite 
wavelength the continua of critical modes associated with second
sound and the dual longitudinal photon  are indistinguishable.
However, at finite momenta they behave differently, tracking the
weight reshuffling found in the quasiparticle spectrum of the
disorder state closely.

\end {abstract}


\maketitle

\section {Introduction}
The notion of Kramers-Wannier duality \cite {Kramers} has been around for a long time in
statistical physics and field theory but one can still wonder if it is scope is fully appreciated.
Especially in  condensed matter physics it appears to be much more than a mathematical
convenience. One can view it instead as a physical `relativity'
principle associated with order. Order and disorder have no objective meaning but just 
depend on the viewpoint of the observer.  For instance, according to the classic 2D
Ising model duality \cite {RSavit}, an observer equipped with
machinery measuring two point correlators
of the order degrees of freedom (the Ising spins) will be convinced that the low temperature
state shows long range order while the high temperature state is just an entropy dominated
featureless entity. On the other hand, the Kramers-Wannier duality demonstrates that the
high temperature state in fact corresponds with an ordered state, a condensate,  formed from  the 
topological excitations (domain walls) associated with  the low temperature order. An experimentalist 
probing the system with a machine sensitive to the domain wall order would be of the opinion that
the {\em high} temperature state is ordered, while the low temperature state is entropy dominated.
When duality is in charge, disorder is order in disguise and one might think that this camouflage
act is perfect. The dual order is carried by the topological excitations of the direct order. In
the continuum limit it takes an infinity of operations involving the order degrees of freedom to probe
the topological excitations and this is beyond the capacity of any machine builder. Henceforth,
it appears that it is fundamentally impossible for an observer which can only employ order degrees
of freedom to directly measure the order associated with  the dual side with the consequence
that the `order-experimentalist' can only perceive dual order as disorder.

Here we will demonstrate that the above is too strong a statement. This `duality censorship'
seems absolute for the special case of the Ising model in 2D. However, in a way its scalar
order parameter structure is too simple. Here we will focus on a more representative 
example: the 3D XY model which might be alternatively interpreted
as the Bose-Hubbard model in 2+1D at zero chemical potential \cite {FisherWGF},
or either as the 
Abelian-Higgs model in 2+1D of high energy physics at $T=0$.
It is characterized by the well-known 
Abelian-Higgs \cite {Kleinert1} or vortex duality which maps the global XY
model on $U(1)/U(1)$ gauge theory. 
In the Bose-Hubbard interpretation, the quantum disordered neutral superfluid corresponds
with a dual 
Meissner phase characterized by Bose condensed vortex-particles.
This incompressible state corresponds
physically with the Bose Mott-insulator \cite {FisherWGF}.

The excitations of the `dual side' are the Higgs (amplitude) mode and massive photons of the dual
 superconductor. Surely, the Higgs mode is subjected to dual censorship for the same reasons as
 found in  the Ising model, but the photons are a different story. As is well known, the Goldstone mode
 (second sound) of the superfluid turns into a doublet of massive excitations in the Mott-insulator 
 corresponding with the propagating unoccupied- and doubly occupied states in the Mott state. As we
 will show here, these actually correspond with linear combinations of the degenerate  transversal- 
 and longitudinal photons of the dual superconductor. The latter is of course the usual extra gauge
 mode associated with the presence of the dual phase order, and one can say that the dual censorship
 does not  prohibit the dual phase order to manifest itself on the order side.  
 
 Our working horse is a simple expression relating the second sound propagators to the dual photons
 which just follows from the Legendre transformation (section III). In combination with the essentially
 complete understanding of the physics resting on the  dual order, we are able to describe with little
 effort some features of the order propagators in the disordered state  which are to the best of our knowledge
 not recognized. On the gaussian level (section IV, V) we obtain the outcome sketched in
 Fig. \ref {FigResponse}b:
 the single Goldstone of the ordered state (Fig. \ref {FigResponse}a) turns into a
 massive doublet in the disordered state.
 At zero momentum the two poles of the velocity propagator are indistinguishable but one of them looses
 its strength when momentum is exceeding the inverse London length of the dual superconductor. This
 makes sense because at shorter lengths order is re-emerging and the simple Goldstone spectrum of
 the ordered state should be recovered.
 
 The Abelian Higgs model in 2+1D is below its upper critical dimension  and its critical state is strongly
 interacting. Resting on the complete description due to Hove and Sudb\o
 \cite {HoveSudbo} of this critical state in the
 dual language, we will derive the critical velocity-velocity propagators (section VI) with the  surprising 
 outcome that its transversal- and longitudinal components  appear to be quite different although
 governed by the same anomalous dimension, again reflecting  the rather different status of the
 'order' (transversal) and 'disorder' (longitudinal) photons when measured through velocity
correlations.  Before getting deeper into this, let us start with some simple considerations
regarding the mode counting.

\section {A simple counting argument}

\begin{figure} 
\includegraphics[width=4.2cm]{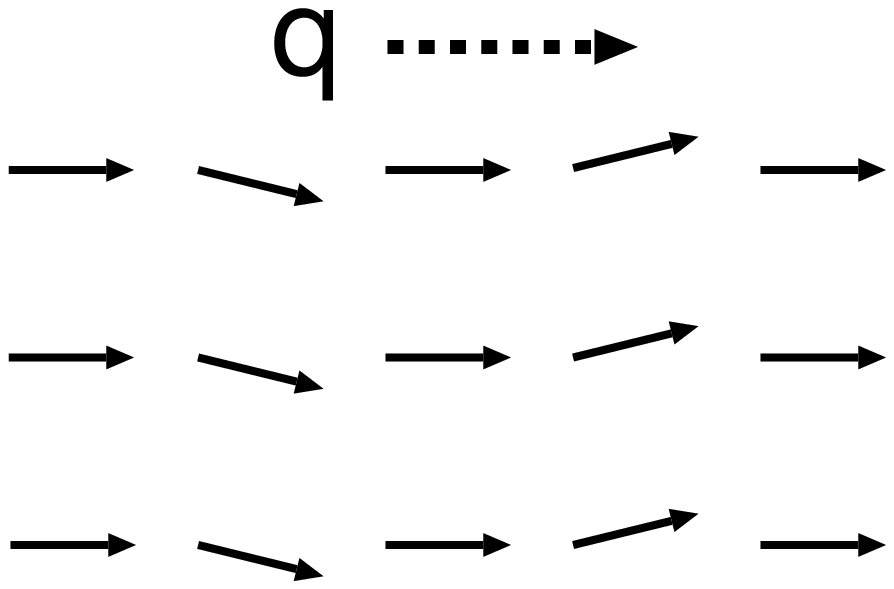}
\includegraphics[width=4.2cm]{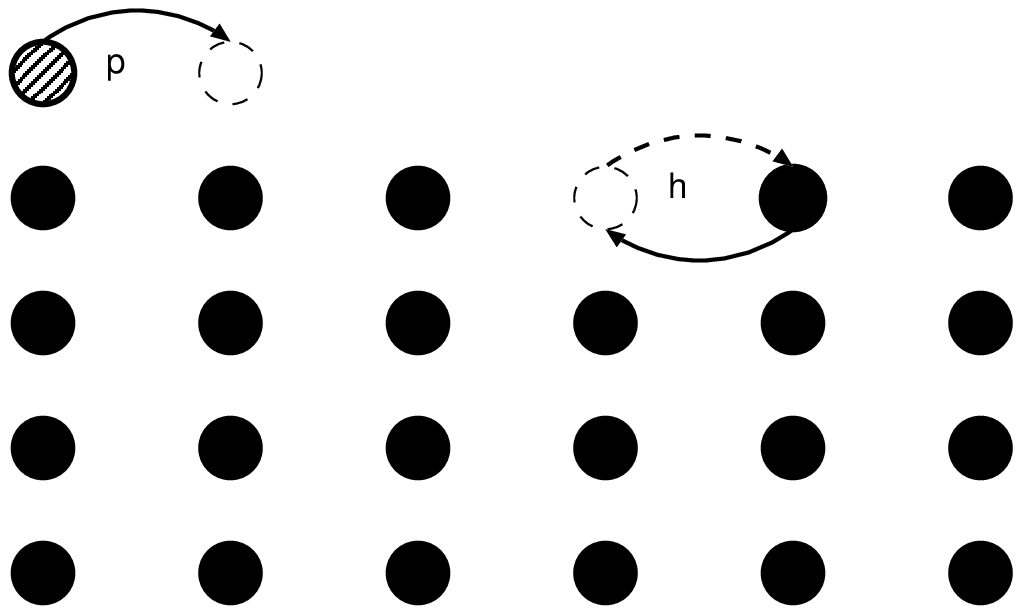}
\begin {center} a) \hspace {4cm} b) \end {center}
\includegraphics[width=4.2cm]{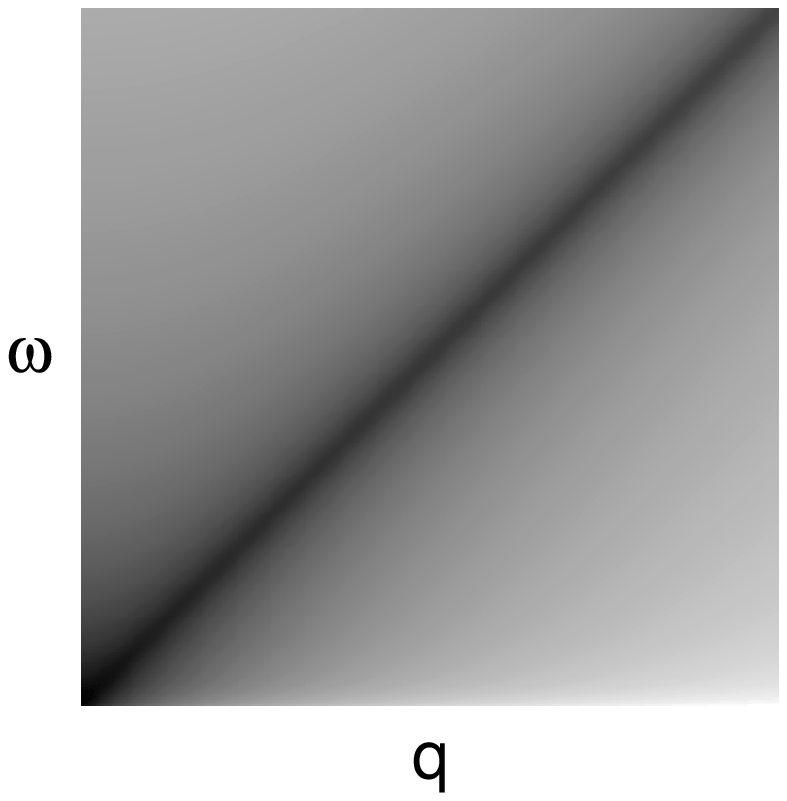}
\includegraphics[width=4.2cm]{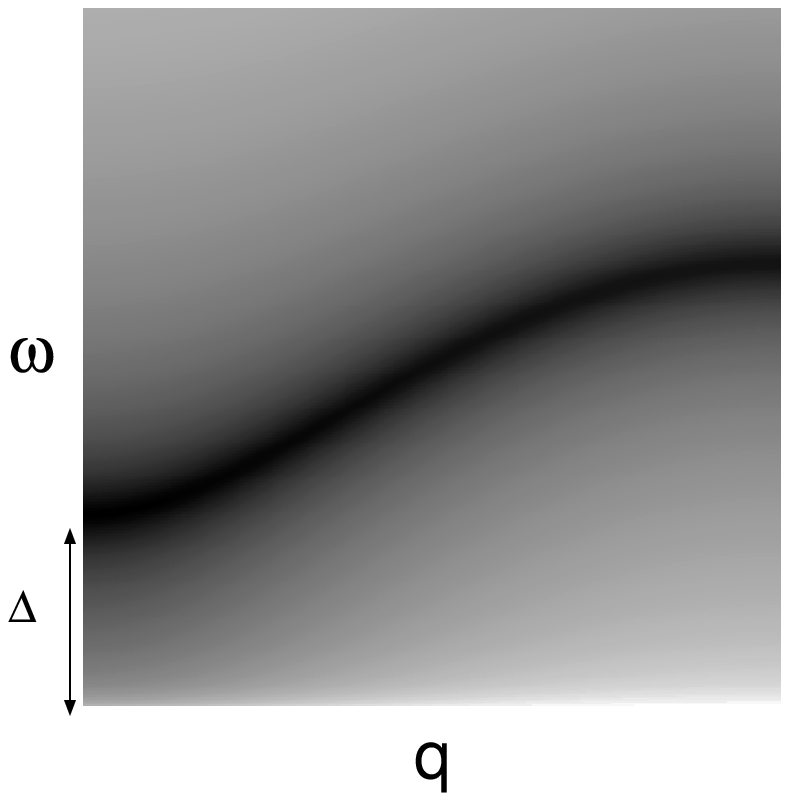}
\begin {center} c) \hspace {4cm} d) \end {center}
\caption{The excitations in the weak/strong coupling limits of the Bose Hubbard model
at zero chemical potential: The Goldstone boson (second sound) with linear dispersion (c)
associated with the superfluid  (phase ordered) state (a) at weak coupling.
In the strong coupling limit (b) a doublet of massive 'excitons' are found with gap $\Delta$ (d)
corresponding with propagating  unoccupied and doubly occupied sites, which 
can be alternatively understood at $q \rightarrow 0$ as the $\pm 1$ angular momentum eigenstates
of a $O(2)$ quantum rotor.} \label {FigCounting}
\end{figure}

The system of interest is the well-known Bose-Hubbard model in 2+1D at vanishing chemical potential, 
written in phase-number representation as\cite {FisherWGF},
\begin{equation}
H = \frac{1}{C} \sum_i n^2_i - J \sum_{\langle i j \rangle} \cos (\phi_i - \phi_j)
\label{hamphase}
\end{equation}
defined on a 2D bipartite lattice; $n_i$ and $\phi_i$ are the number- and phase operators on
site $i$, satisfying the commutation relation $[ n_i, \phi_j ] = i \delta_{ij}$. The
first- and second term in Eq. (\ref{hamphase}) represents the charging- and Josephson energy
respectively. When the coupling constant $\tilde g = 2/ (JC)$ is small the Josephson energy will
dominate and the phase is ordered at zero temperature, while the excitation spectrum consists
of a single Goldstone mode (phase mode or second sound) shown in Fig.
\ref {FigCounting}a, c. On the other hand, when $\tilde g$ is large the phase is quantum
disordered and number condenses such that $n_i = 0$ modulo local fluctuations, signaling
the Mott-insulator. Of central interest is the excitation spectrum of the Mott-insulator.
In the rotor language \cite {SubirQPT}, the ground state is the angular momentum singlet
while the lowest lying
excitations consist of a doublet of propagating $M = \pm 1$  modes characterized by a
zero-momentum mass gap (Fig. \ref {FigCounting}d). In the Bose-Hubbard interpretation these have a simple 
interpretation in the strong coupling limit   ($\tilde g \rightarrow \infty$) as bosons added 
($M= +1$) or removed ($M = -1$) from the
charge-commensurate state (Fig. \ref {FigCounting}b), while their delocalization in the
lattice produces an identical dispersion
due to the charge conjugation symmetry of the model Eq. (\ref{hamphase}).

This is of course overly well known, but one might be at first a bit puzzled about how to relate
these excitations to the well established notion that the phase disordered state does
correspond with a dual (vortex) superconductor. After all, the doublet of massive modes are fully
protected elementary excitations and they should arise regardless the way one wants to describe the
system.

In the scaling limit, sufficiently close to the quantum phase transition where a continuum 
field-theoretic description applies, phase dynamics might as well be described in terms
of the dual 'disorder field theory'.  This turns out to be just the Ginzburg-Landau-Wilson
theory of a relativistic 2+1D  $U(1)$ superconductor \cite {Kleinert1}, 
\be
  {\mathscr L}_{EM, full} &=& \tfrac {\tilde g} 4 F_{\mu \nu} F_{\mu \nu} + \tfrac 1 2 | (\partial_\mu - i A_\mu) \Psi_V |^2 +  \nonumber \\
  && \tfrac 1 2 m^2 |\Psi_V|^2 + \omega |\Psi_V|^4 \label {L_EM_full}
\ee

The Higgs field $\Psi_V$ describes the Bose condensate of vortices, i.e. the tangle of vortex 
worldlines. In 2+1D vortex 'particles' are in a precise sense indistinguishable from electromagnetically
charged particles. The long range interactions between vortices mediated by the phase 
condensate can be described in terms of non-compact $U(1)$ gauge fields $A_{\mu}$, $U(1)$
appearing in the
connections for the matter field and in the field strength $F_{\mu \nu} = \partial_{\mu} A_{\nu} -
\partial_{\nu} A_{\mu}$. Of crucial importance for what comes, Eq.(\ref{L_EM_full}) is
{\em fully relativistic}. The phase-dynamics problem is characterized by a single (spin-wave)
velocity which we take to be one. Although in dual representation we are dealing with two
separate fields $\Psi_V$ and $A_{\mu}$, both matter- and gauge fields are governed by the same velocity.

The ordered- and disordered phase correspond with the Coulomb- and  Meissner phase,
respectively,  of the theory Eq. (\ref{L_EM_full}).  Obviously, the observable consequences
of the dual theory Eq. (\ref{L_EM_full}) and the original theory  Eq. (\ref{hamphase}) have to
be the same  and for this to be the case, it is a necessary condition
that the mode content of both theories are the same.

Let us first consider the superfluid phase/dual Coulomb phase. Since phase is condensed,
the system should be characterized by a {\em single} Goldstone boson (spin-wave/second sound). 
How to count this on the dual side? The dual Lagrangian is just the 2+1D 
non-compact U(1) Maxwell Lagrangian $F_{\mu \nu} F^{\mu \nu}$. This is characterized
by three vector potentials $A_{\tau}, A_{x}, A_{y}$ and one gauge constraint. Accordingly,
the dual theory is characterized by {\em two} physical degrees of freedom. This is confusing
at first, but as we will  discuss in some detail in the next section, it makes a perfect sense.
One photon  is dynamical and transverse ( $A_T = \frac {\nabla \times {\bf A}}{\nabla}$)
and is just the Goldstone mode in the dual language.
The other physical photon is the temporal one $A_\tau$ ,
describing the instantaneous ('Coulomb') interaction between static vortex
sources: in the Coulomb gauge the longitudinal photon ($A_L = \frac {\nabla \cdot {\bf A}}{\nabla}$) drops
out as we subject it to the Coulomb gauge fix $\nabla \cdot {\bf A} \equiv 0$. Hence, we find 
the correct mode content: the gauge field description is just more efficient than the phase description,
because the instantaneous vortex-vortex interactions appear explicitly, while they have to be
constructed by hand in terms of the phase degrees of freedom.

The full powers of the dual gauge theory unfold in the phase-disordered state. In terms of phase,
all one has is a strong coupling expansion in the Hamiltonian language of Eq. (\ref{hamphase})
we already alluded to, leading to the conclusion that at least in the long wavelength limit
one is dealing with a twofold degenerate massive 'exciton'. What is the dual gauge theory
telling us? As will become clear later, the vortex amplitude 'Higgs boson' $|\Psi_V|$ is subject
of dual censorship, and what remains are the original photons $A_{\mu}$
and the dual phase degree of freedom $\phi_V$ defined through $\Psi_V = |\Psi_V| e^{i\phi_V}$.
If the system would be non-relativistic such that $c_V / c \rightarrow \infty$, where $c_V$ and $c$
are the vortex condensate phase velocity and spin wave velocity, respectively, one would run
into a problem. In this limit,  the vortex phase drops out as a dynamical degree of freedom, and
the photons are counted in the same way as in the Coulomb phase except that they are now
massive: the interactions between static vortices are now screened, while the transversal
'spin wave' photon acquires a mass. In this way one would find {\em one}
propagating massive mode instead of two.

In the relativistic theory ($c_V = c$) it fits neatly: we encounter now four fields ($A_{\tau}, A_x,
A_y, \phi_V$) subjected to one gauge constraint. Of the remaining three physical fields one
takes care of static interactions, and we are left with two photons which are degenerate! 
Since only one of these photons is carried by the smooth phase field configurations, the other 
one reflects the phase rigidity of the vortex condensate:
one of the two excitons lighting up in the phase world is just telling that the dual superconductor
resists a  twist in its phase.

Having given away the bottomline, let us now substantiate these matters with some explicit
computations.

\section{Disorder fields probed by order means}   

It does not seem to  be widely recognized that the propagators of order fields can be
straightforwardly expressed  in terms of the disorder fields. This was discussed in
an earlier paper dealing with field-theoretical elasticity \cite {ZMN}, and to save the reader the
effort of learning this intricate affair, let us rederive these relations for the far simpler
Abelian-Higgs case. 

These relations are associated with the first step of the duality, where the phase fields
are turned into  photons mediating interactions between the vortices. To
set the stage, let us shortly review these matters\cite {ZeeDuality, DasguptaHalperin, Kleinert1, DHLeeMPAFisher, FazioZant, TBanks,
NguyenSudbo, HoveSudbo, SenthilFisher, Tesanovic, YBCO}.
After coarse graining,  phase dynamics can be written in terms of the Lagrangian,
\be
  {\mathscr L}_{XY} = \frac 1{2 g} \left \lbrack (\partial_\tau \phi)^2 + c^2 (\nabla \phi)^2 \right \rbrack \rightarrow  \frac 1{2 g} (\partial_\mu \phi)^2. \label {L_XY}
\ee
The coupling constant $g$ is proportional to the original coupling constant $\tilde g$.
The spin-wave velocity 
is given by ratio of the stiffness and compression moduli $c^2 = \rho_s / \kappa_s$
and set to 1 in the last step. It is left implicit that the phase field
is compact, $\phi = \phi + 2 \pi$.  We take the superfluid 'three' velocity 
\begin{equation}
v_{\mu} ( {\bf x} ) = \partial_{\mu} \phi ({\bf x}) \label{velocity}
\end{equation}
as the natural, 'primitive' observable of the orderly side. In the ordered
phase, the momentum space velocity-velocity propagator is proportional to the phase-phase propagator,
 $\langle \langle v_{\mu} | v_{\nu} \rangle  \rangle_{q, \omega} = q_{\nu}^2 \delta_{\mu \nu} $ 
 $\langle \langle  \phi | \phi  \rangle  \rangle_{q, \omega}$ and the latter suffices to  calculate the
 order parameter propagator $\langle \langle e^{i\phi} | e^{i\phi} \rangle  \rangle$
 (e.g., ref. \cite{FisherWGF}). In the disordered phase $\phi$ itself becomes
 multi-valued and meaningless, but $v_{\mu}$ continues
 to be single valued and meaningful.
 
In  the phase-ordered state the theory Eq. (\ref{L_XY}) is Gaussian and the velocity
 propagator is easily computed by adding a generating functional to the Lagrangian,
 \be
  {\mathscr L} \lbrack {\cal J}_\mu \rbrack = {\mathscr L}_{XY} + {\cal J}_\mu \partial_\mu \phi, \label {generating_functional_XY}
\ee
followed by taking the functional derivative
\be
  \langle \langle v_\mu | v_\nu  \rangle \rangle = \frac 1 Z \left . \frac {\partial^2 Z \lbrack {\cal J}_\mu \rbrack}{\partial {\cal J}_\mu \partial {\cal J}_\nu} \right |_{{\cal J}_\mu = 0}. \label {phi_propagator_relativistic}
\ee
The non-relativistic propagator measured in condensed matter experiments
represents only the subset of components of the relativistic propagator Eq. (\ref {phi_propagator_relativistic}) with spatial indices: $\langle \langle v_i | v_j  \rangle \rangle$. In the phase ordered state of the XY model  one can integrate the Gaussian Goldstone
fields in Eq. (\ref{generating_functional_XY}) with the result,
\be
  Z \lbrack {\cal J}_\mu \rbrack = \prod_{p_\mu} \sqrt {\frac {2 \pi g}{p^2}} ~ e^{\tfrac g 2 {\cal J}_\mu \frac {p_\mu p_\nu}{p^2} {\cal J}_\nu}. \label {Z_XY_J_smooth}
\ee
and the propagators follows immediately from Eq. (\ref{generating_functional_XY}) identity.
The relativistic and non-relativistic versions are respectively,
\begin{eqnarray}
  \langle \langle v_\mu  | v_\nu  \rangle \rangle & = & g 
\frac {p_\mu p_\nu}{p^2}, \label{propagator_XY_relativistic} \\
 \langle \langle v_i  | v_j  \rangle \rangle & = & g 
\frac { c^2 q^2}{\omega_n^2 + c^2 q^2} P^L_{ij}. \label{propagator_XY_nonrelativistic}
\end{eqnarray}
where the longitudinal- and transversal (for later use) projection operators are,
\be
  P_{ij}^L = \frac {q_i q_j}{q^2}, \qquad
  P_{ij}^T = \delta_{ij} - \frac {q_i q_j}{q^2}. \label {PLPT}
\ee
Obviously, we find only a single mode: the single Goldstone boson of the scalar theory as shown
response plot Fig. \ref {FigResponse}a.

Of course, the above procedure no longer works in the absence of the phase condensate.
At any finite disorder there are configurations present  containing topological defects (vortices) ,
which are ignored in the path integral Eq. (\ref {Z_XY_J_smooth}) but these are easily handled 
in the language of the dual disorder field theory. 

Let us turn to the duality itself. The first crucial step is  a simple Legendre transformation.
Introduce Hubbard-Stratanovich auxiliary fields $\xi_{\mu}$ such that the phase action
including the external sources, Eq.(\ref{generating_functional_XY}) is written as,
\be
{\mathscr L} \lbrack {\cal J}_\mu \rbrack = 
- \tfrac g 2 {\cal J}_\mu {\cal J}_\mu + i g {\cal J}_\mu \xi_\mu + \tfrac g 2 \xi_\mu \xi_\mu + i \xi_\mu \partial_\mu \phi \label{hubstra}
\ee
Let us ignore the external sources ${\cal J}_{\mu}$ for a moment, to focus on the duality
itself. The phase field is split into smooth- and multivalued pieces: $\phi= \phi_{sm} +
\phi_{MV}$. By reshuffling derivatives ($\xi_{\mu} \partial_{\mu} \phi_{sm} = - \phi_{sm}
\partial_{\mu} \xi_{\mu}$) the Gaussian $\phi_{sm}$ turn into Lagrange multipliers
imposing a conservation law on the auxiliary fields,
\be 
\partial_{\mu} \xi_{\mu} = 0 \label{consxi}
\ee
In  the phase dynamics interpretation, $\xi_{\mu}$ just represents the supercurrents and Eq. (\ref{consxi})
is the hydrodynamical  continuity equation governing superflow. In 2+1D continuity can be
imposed on the supercurrents by writing $\xi_{\mu}$ in terms of non-compact $U(1)$ gauge 
fields $A_{\mu}$ as
\be
\xi_{\mu}=\varepsilon_{\mu \nu \lambda} \partial_{\nu} A_{\lambda} \label{intrA}
\ee
and it follows that the remaining pieces of the  Lagrangian Eq. (\ref{hubstra}) (ignoring the ${\cal J}$'s) 
can be written as,
\be
\frac{g}{2} \xi_{\mu} \xi_{\mu} + i \xi_{\mu} \partial_{\mu} \phi_{MV} =  \tfrac {g}{4} F_{\mu \nu} F^{\mu \nu}
+ i A_{\mu} J^V_{\mu} \label{dualac}
\ee
where $F_{\mu \nu} = \partial_{\mu} A_{\nu} - \partial_{\nu} A_{\mu}$ is the field strength of the
dual gauge sector while 
\be
J^V_{\mu} = \varepsilon_{\mu \nu \lambda} \partial_{\nu} \partial_{\lambda} \phi_{MV} \label{vortexcur}
\ee
is the non-integrability of the phase field, having the meaning of vortex current. This completes
the key step of the duality: it shows that in 2+1D the rigidity of the phase medium can be
exactly parameterized in `force carrying photons' (the $A_{\mu}$'s) while the vortices act as
sources for these photons. The disorder field theory Eq. (\ref{L_EM_full}) follows immediately:
vortices are indistinguishable from charged bosons interacting with electromagnetic fields and
a system of such bosons is described by the Ginzburg-Landau-Wilson theory \cite {Kleinert1}.

This is all familiar territory but the following simple operation seems not to be commonly
known.   Let us include the external sources and use the identity Eq. 
(\ref{phi_propagator_relativistic}) to calculate the phase propagator, but now using the
action after the Hubbard-Stratanovich transformation,     
\begin {widetext}
\be
  \left . \frac {\partial^2 Z \lbrack {\cal J}_\mu \rbrack}{\partial {\cal J}_\mu \partial {\cal J}_\nu} \right |_{{\cal J}_\mu = 0} &=& \int \pathD \xi_\mu \pathD \phi ~ \left . \frac {\partial^2}{\partial {\cal J}_\mu \partial {\cal J}_\nu} e^{\int \rmd x_\nu ~ \left ( \tfrac g 2 {\cal J}_\mu {\cal J}_\mu - i g {\cal J}_\mu \xi_\mu - \tfrac g 2 \xi_\mu \xi_\mu - i \xi_\mu \partial_\mu \phi \right )} \right |_{{\cal J}_\mu = 0} \nonumber \\
  &=& \int \pathD \xi_\mu \pathD \phi ~ (g \delta_{\mu \nu} - g^2 \xi_\mu \xi_\mu) ~ e^{\int \rmd x_\nu \left (- \tfrac g 2 \xi_\mu \xi_\mu - i \xi_\mu \partial_\mu \phi \right )} 
  = Z \left ( g \delta_{\mu \nu} - g^2 \langle \langle \xi_\mu | \xi_\nu \rangle \rangle \right )
\ee
\end {widetext}
This implies an exact relationship between the velocity propagator and the propagator of 
the supercurrents,  rooted in the Legendre transformation,
\be
  \langle \langle v_\mu  | v_\nu  \rangle \rangle = g \delta_{\mu \nu} - g^2 \langle \langle \xi_\mu | \xi_\nu \rangle \rangle. \label {Zaanen_Mukhin}
\ee
Because $\xi_{\mu} = \varepsilon_{\mu \nu \lambda} \partial_{\nu} A_{\lambda}$ this implies that
in fact the phase velocity/spin wave propagator is proportional to a linear combination of 
the physical photon propagators  of the dual gauge disorder-field
theory. This implies that the poles of the magnon and photon propagator have to coincide and
this has to be because both describe the same physics. However, the pole strengths
might be quite different reflecting the `dual relativity principle': pending the use of order or
disorder `tools' one might get a very different view  of the same underlying reality. The result
Eq.(\ref{Zaanen_Mukhin}), first derived in Ref. \cite{ZMN}, shows that at least in the Abelian-Higgs
case, the two observers should actually agree more on what they see than  one could have expected a-priori.
The key is that although the vortex condensate falls prey to dual censorship, the orderly
observer can still learn much about the dual world because he/she can probe the dual 
photons according to Eq.(\ref {Zaanen_Mukhin}).

\section {Magnons as photons}

Let us exercise the notions of the previous section in the simple case of the phase ordered state.
We know the answer (the Goldstone mode, Eq. \ref{propagator_XY_relativistic}), we know the dual
side (Maxwell theory, $F_{\mu \nu} F^{\mu \nu})$, and we know how these relate (the
`Zaanen-Mukhin' relation, Eq. \ref{Zaanen_Mukhin}). It is indeed a straightforward exercise. 
 
Although the dynamics is fully relativistic the questions of relevance to condensed matter
experimentalists are not relativistic: only the spatial components of the propagator are 
measurable (Eq. \ref{propagator_XY_nonrelativistic}).  This makes it convenient to
use a Coulomb gauge fix. The Maxwell action in momentum-Matsubara frequency
space becomes, including the external sources $J_\mu^{ext.}$,
\be
  {\mathscr L}_{EM} &=& \tfrac g 2 \left (  A^\dagger_\tau ,  {\bf A}^\dagger  \right ) \left ( \begin {array}{c c} q^2 & -i \omega_n \langle {\bf q} | \\ i \omega_n | {\bf q} \rangle & \omega_n^2 \hat {\mathbf 1} + c^2 q^2 \hat P^T \end {array} \right ) \left ( \begin {array}{c} A_\tau \\ {\bf A} \end {array} \right ) \nonumber \\
  && + i J_\tau^{ext.} A_\tau + i {\bf J}^{ext.} \cdot {\bf A}^\dagger.
\ee
where we have explicitly indicated the time- ($X_{\tau}$) and space (${\bf X}$) components of
the gauge fields and currents. The bra- and ket in the gauge field propagator represent rows and columns 
$q_i$, respectively. Notice that from now on we keep the Goldstone/spin-wave velocity $c$
explicit, for purposes which will become clear later.
 
Provided that we choose a gauge fix ${\cal F}$ that does not act 
on the temporal component $A_\tau$, the temporal component can be integrated out.  This yields
the usual Lagrangian with Coulomb interactions between  static sources, 
\be
  {\mathscr L}_{EM} &=& \tfrac 1{2g} \frac {J^\dagger_\tau J_\tau}{q^2} + \tfrac g 2 \left (\omega_n^2 + c_{}^2 q^2 \right ) {\bf A}^\dagger \hat P^T {\bf A} + \nonumber \\
   && + i (J_L - \frac {i \omega} q J_\tau) A^\dagger_L + i {\bf J} \hat P^T {\bf A}^\dagger. \label {L_EM_Coulomb}
\ee
The longitudinal 
component $A_L$ is unphysical (its source is $i \omega_n J_\tau - q J_L \rightarrow
\partial_\tau J_\tau + \partial_i J_i =0$) and it should be removed 
by the Coulomb gauge 
\be
  0 = \partial_i A_i = q A_L. \label {Coulomb_gauge_fix}
\ee
We end up with two propagators for the gauge fields, as it should in 2+1D. We find one dynamical
photon,  
\be
  \langle \langle A^\dagger_i | A_j \rangle \rangle = \frac {P^T_{ij}}{g (\omega_n^2 + c^2 q^2)}. \label {propagator_Aij}
\ee
and a propagator taking care of the Coulomb interactions between the static sources,
\be
  \langle \langle A^\dagger_\tau | A_\tau \rangle \rangle = \frac 1 {g q^2} \label {propagator_Atau}
\ee
This is of course textbook electromagnetism, but be aware of the twist in the interpretation. The
photons now keep track of the capacity of the phase condensate to respond to external influences.
The outcome is: it is carrying a 'Goldstone photon' ($A_T$, Eq. \ref{propagator_Aij}) and it
can mediate as well
interactions between static vortices ($A_{\tau}$, Eq. \ref{propagator_Atau}).

We are now in the position to evaluate  the 'Zaanen-Mukhin' relation Eq. (\ref{Zaanen_Mukhin}). 
For this purpose, we are only interested in the spatial components of  the supercurrents $\xi_{\mu}$.
The supercurrent propagator is easily found by using the definition Eq. (\ref {intrA}), and the results
for the gauge field propagators Eqs. (\ref {propagator_Aij}, \ref {propagator_Atau}) and we find for
its spatial components,  
\be
  \langle \langle \xi^\dagger_i | \xi_j \rangle \rangle = \frac 1 g \left \lbrack \frac {\omega_n^2}{\omega_n^2 + c_{ph}^2 q^2} P^L_{ij} + P^T_{ij} \right \rbrack.
\ee
Using now the Zaanen-Mukhin relation Eq. (\ref{Zaanen_Mukhin}),
\begin{eqnarray} 
\langle \langle v_i  | v_j  \rangle \rangle & = & g \delta_{i j} - g^2 \langle \langle \xi_i | \xi_j \rangle \rangle \nonumber \\
& = & g \lbrack  P^L_{ij} + P^T_{ij}  \rbrack - g \left \lbrack \frac {\omega_n^2}{\omega_n^2 + c_{ph}^2 q^2} P^L_{ij} + P^T_{ij} \right \rbrack \nonumber \\
& = & g \frac{c^2 q^2}{\omega_n^2 + c^2 q^2} P^L_{ij}.
\label{goldres}
\end{eqnarray}
After this long detour, we indeed have managed to recover the spin wave propagator
Eq. (\ref {propagator_XY_nonrelativistic})! 

The simple lesson following from this simple exercise is that the dual photon language
is in a way more complete than the description in terms  of phase fields, in the sense that the gauge
fields keep track in an explicit way of both the capacity of the medium to propagate
Goldstone bosons and the fact that it mediates interactions between its topological
excitations. The Zaanen-Mukhin relation filters out the Goldstone sector from the
'omnipotent' dual gauge sector, keeping its topological side (the Coulomb propagator, requiring
vortex sources) completely hidden from the eye from the 'orderly' observer.

\section {Watching the disordered state with orderly means}

Surely, the dual route of the previous section  is a rather inefficient
way to derive the propagator of a Goldstone mode. This changes drastically in the
phase disordered state. Resting on the fact that the dual gauge theory is now governed
by order, precise information on the second sound
propagator can be extracted  with barely any extra investments.  The only other option
is the strong coupling expansion in the Hamiltonian language and this  becomes
very tedious at intermediate couplings. 

The phase disordered state corresponds with the Higgs phase of the gauge theory  
 Eq.(\ref{L_EM_full}), corresponding with the state where vortex loops have blown
 out and the vortices have Bose-condensed.
As a consequence, the bosonic disorder field $\Psi = |\Psi_0| e^{i \phi_V}$ acquires a finite 
expectation value. This theory is fully relativistic, as we explained, and this vortex condensate
is literally like the $U(1)$ Higgs phase of high energy physics \cite {Higgs}.  It will turn out
to be quite convenient for the interpretation of the results to consider a non-relativistic
extension of the theory characterized by a condensate velocity $c_V$, which is different
from the spin-wave velocity $c$, entering the time components of the covariant derivatives
$ \sim  \frac {1}{c_V} ( \partial_{\tau} - iA_{\tau} )$. Of course, the real theory is characterized
by $c_V = c$ as implied by the Lorentz-invariance of the action Eq. (\ref {L_XY}).

Let us employ the usual unitary gauge, corresponding with fixing the condensate phase $\phi_V = 0$.
 The finite expectation value of the disorder field results in the familiar Higgs term in the action
\be
  {\mathscr L}_{Higgs} = \tfrac 1 2 |\Psi_0|^2 \left \lbrack \tfrac 1{c_V^2} A_\tau A_\tau + A_i A_i \right \rbrack. \label {L_Higgs_unitary}
\ee
The only specialty is the velocity $c_V$. In high energy physics this is the light velocity while
in the non-relativistic condensates of condensed matter physics $c_V$ is the sound velocity
(in BCS theory $\sim v_F$  \cite {AndersonHiggs,Galitskii}),
which is vanishingly small compared to the light velocity with the consequence that one can get
away with a time independent Ginzburg-Landau theory. In our final result we have to set $c_V = c$.

We now follow the same route as in the previous section. Adding the Higgs term
the Lagrangian becomes,
\begin {widetext}
\be
  {\mathscr L}_{full} &=& \tfrac g 2 \left(  A^{\dagger}_\tau , {\bf A}^{\dagger} \right ) \left ( \begin {array}{c c} q^2 + \frac {\Omega^2}{c_V^2} & - \omega_n \langle {\bf q} | \\ - \omega_n | {\bf q} \rangle & ( \omega_n^2 + \Omega^2 ) \hat {\mathbf 1} + c^2 q^2 \hat P^T \end {array} \right ) \left ( \begin {array}{c} A_\tau \\ {\bf A} \end {array} \right ) + i {\cal J}_\tau A^\dagger_\tau + i {\bf J} \cdot {\bf A}^\dagger. \label {L_EM_Higgs_full}
\ee
\end {widetext}
Introducing a Higgs mass $\Omega$, defined by
\be
  \Omega^2 = \frac {|\Psi_0|^2}g. \label {Omega}
\ee

Since the gauge has already been fixed, the temporal components $A_\tau$ can be safely
integrated out,
\begin {widetext}
\be
  {\mathscr L}_{full} &=& \tfrac g 2 {\bf A}^\dagger \left \lbrack \frac {\Omega^2 (\omega^2 + c_V^2 q^2 + \Omega^2)}{c_V^2 q^2 + \Omega^2} \hat P^L + (\omega_n^2 + c^2 q^2 + \Omega^2) \hat P^T \right \rbrack {\bf A} + i {\cal {\bf J}} \left \lbrack \hat {\mathbf 1} - \frac {c_V^2 q^2}{c_V^2 q^2 + \Omega^2} \hat P^L \right \rbrack {\bf A}^\dagger + \tfrac 1{2g} \frac {J^\dagger_\tau J_\tau}{q^2 + \frac {\Omega^2}{c_V^2}}. \label {L_EM_unitary}
\ee
\end {widetext}

The last term corresponds with the interactions between the static vortices which are now short ranged.
The interest is in the dynamics of the gauge fields itself. As before, we find a transversal
photon $A_T$ characterized by  second sound propagator which has acquired a Higgs mass.
In addition, we find an extra longitudinal photon (the first term) which is now physical.
This is also characterized by the same Higgs mass but it is propagating at the condensate velocity
showing that it represents the phase rigidity of the dual superconducting matter sector.

The propagators for gauge fields are easily determined from the inverse of the full action Eq. (\ref {L_EM_Higgs_full}). The superfluid current propagator is decomposed into longitudinal and transversal
parts $\xi^{L, T}$ (parallel and perpendicular to the momentum ${\bf q}$ respectively) and
the  propagators are found to be
\be
  \langle \langle \xi_L | \xi_L \rangle \rangle &=& \tfrac 1 g \frac {\omega_n^2}{\omega_n^2 + c_{}^2 q^2 + \Omega^2}, \label {xiL_Green_disorder} \\
  \langle \langle \xi_T | \xi_T \rangle \rangle &=& \tfrac 1 g \frac {\omega_n^2 + c_V^2 q^2}{\omega_n^2 + c_V^2 q^2 + \Omega^2}. \label {xiT_Green_disorder}
\ee

\begin{figure} 
\includegraphics [width=4.2cm]{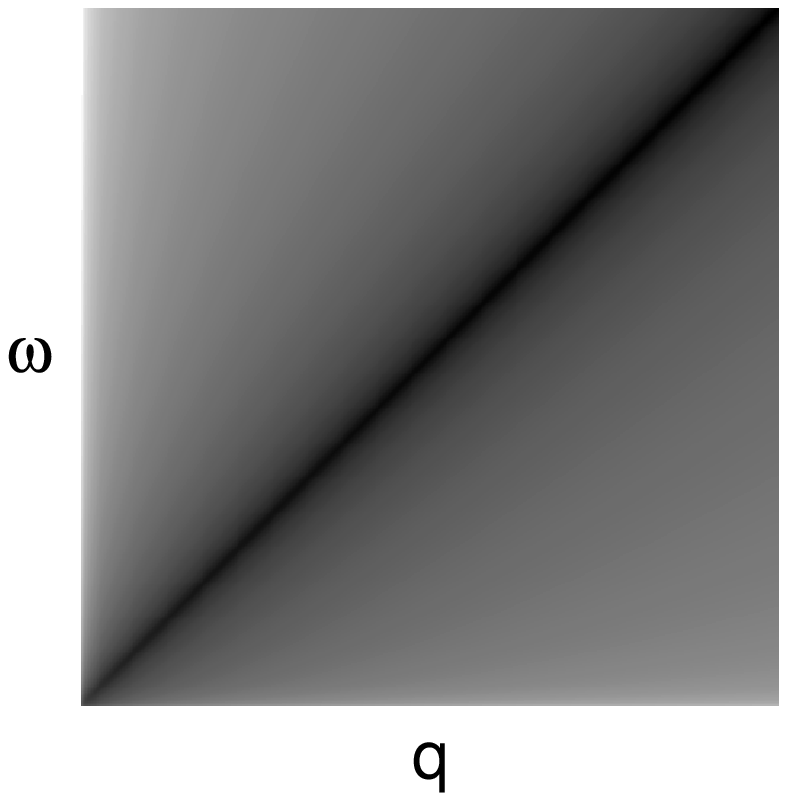}
\includegraphics [width=4.2cm]{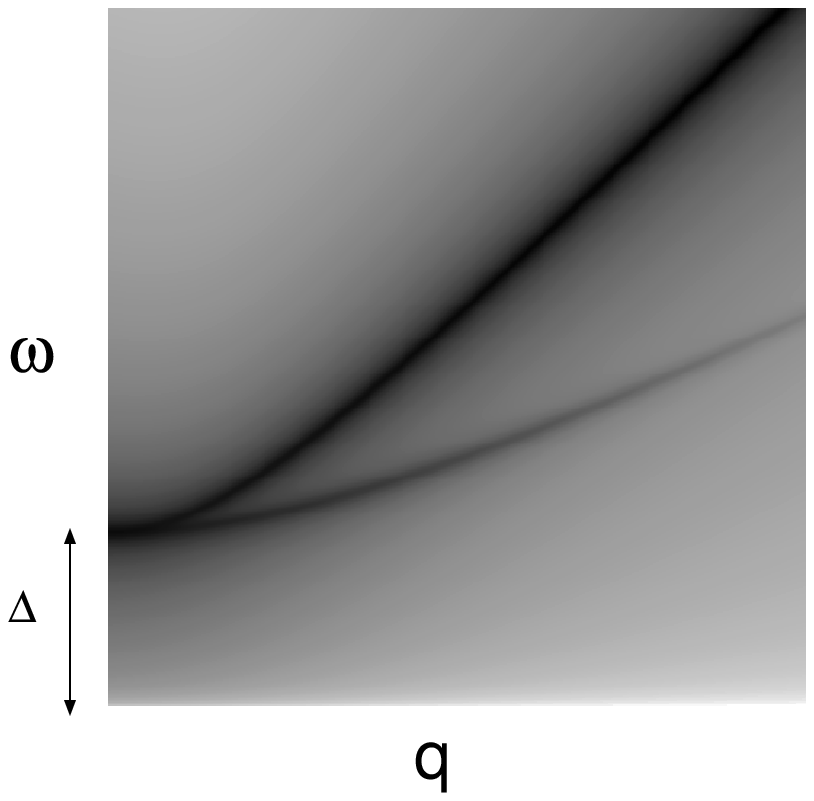}
\begin {center} a) \hspace {4cm} b) \end {center}
\caption{Spectral functions associated with the superfluid velocity-propagator,
computed on the Gaussian level using the dual theory . These results should
be accurate deep inside the ordered and disordered phase
(a) The ordered (superfluid) phase: the second sound pole of Fig. \ref {FigCounting}c is recovered as 
it should. (b) The disordered (Mott insulating) phase: we used here a condensate velocity ($c_V$)
which is half the sound velocity $c$ for the mere purpose to make visible  the different behaviors
of the strength of the second sound- (higher branch) and dual (vortex) condensate (lower branch)
poles. In reality these velocities are the same and the modes are degenerate. For $q \rightarrow 0$
the pole strengths  of the  two modes are the same, while they are governed by the same Higgs mass,
and they can be combined in the $\pm 1$ helicity modes as expected from the strong coupling
expansion in the Hamiltonian formalism (Fig. 1d; see section V). However, for increasing momentum the
condensate pole looses gradually strength while the second sound pole becomes more and more like
the 'orderly' result of (a), reflecting that at distances short compared to the dual London penetration
depth the medium 'rediscovers' the order.}   
 \label {FigResponse}
\end{figure}

Using now the Zaanen-Mukhin relation
Eq. (\ref {Zaanen_Mukhin}) and the momentum propagators Eqs. (\ref {xiL_Green_disorder} -
\ref {xiT_Green_disorder}), we obtain the result for the non-relativistic propagator for the superfluid velocity in
the disordered phase,
\be
  \langle \langle v_i  | v_j  \rangle \rangle = g \left \lbrack \tfrac {c^2 q^2 + \Omega^2}{\omega_n^2 + c^2 q^2 + \Omega^2} P^L_{ij} + \tfrac {\Omega^2}{\omega_n^2 + c_V^2 q^2 + \Omega^2} P^T_{ij} \right \rbrack
 \label{mottprop} 
\ee
The spectral response from this propagator is plotted on Fig. \ref {FigResponse}b.

The longitudinal (first) term represents, as before ( Eq. \ref {goldres}), the correlations associated with
the smooth part of the phase field : this is literally second sound acquiring a mass
associated with the disappearance of the superfluid rigidity  at large lengths- and
times.   We notice that in the static limit ($\omega_n \to 0$) the longitudinal part becomes
a constant, signaling that even at the shortest distances superfluid correlations have disappeared.
This makes sense: when vortices populate the whole system, then any long living correlation is 
destroyed even between two neighboring sites when one waits long enough.

The second, transversal term is the interesting one: we indeed find a second mode and although
it has the same mass as the gapped second sound it propagates with the condensate velocity. It
is of course the longitudinal photon reflecting the dynamics of the dual superconducting vortex
matter. In order for the superfluid velocity correlator to acquire a non-zero transversal component
it is actually a requirement that the phase field becomes non-integrable.  This becomes clear
by inspecting the transversal part of the supercurrent Eq. (\ref {xiT_Green_disorder}) ,
\be
  \xi_T &=& -i e^T_i \xi^i = - e^T_i \frac {\partial_i (\phi_{sm} + \phi_{MV})}g \label {xi_T}  \\
  &=& - \tfrac 1 g e^T_i \left ( i q e^L_i \phi_{sm} + \partial_i \phi_{MV} \right ) = - \tfrac i g e^T_i \partial_i \phi_{MV} \nonumber
\ee
where the smooth part has disappeared since it makes no sense to have derivatives in the transversal
direction for the case of smooth fields.

Let us analyze the result Eq. (\ref{mottprop}) in more detail. The velocity $c_V$ has done its job in
establishing that the transversal poles of Eq. (\ref{mottprop}) are indeed due to the superconducting
vortex matter, and we can impose  Lorentz invariance by setting $c_V = c$: the longitudinal- and
transversal modes become degenerate as they should because of the degeneracy associated with
the $n = \pm 1$ excitons, trivially seen in the Hamiltonian formalism.  A different issue is the
pole strength of the second sound ($I_L$) and condensate poles ($I_T$) as measured by the 
velocity-velocity correlator, characterized by the ratio,
\begin{equation}
\frac {I_L} {I_T} = 1 + \frac{ c^2 q^2} {\Omega^2}
\label{ratio}
\end{equation}
Giving this a minute of thought this makes a perfect sense. First, at $q=0$ both modes at
real frequency $\omega = \Omega$ have the same weight. It follows from the Hamiltonian
formalism that at wavelengths large compared to the London length the excitons correspond
with the helicity $\pm 1$ eigenstates of the angular momentum of the O(2) quantum rotors.
The supercurrents have the status of canonical momenta and should therefore be combined
in currents with definite helicity $\xi_{\pm 1} \sim \xi_L \pm i \xi_T$. The implications is
obvious: at $q \rightarrow 0$ the longitudinal- and transversal poles of the velocity propagator
should have the same  strength because all what exist in this limit are the helicity eigenstates.

What is changing at smaller distances? The characteristic momentum scale is of course
the inverse dual London penetration depth $ q_{L} \simeq 1 / \lambda_L = \Omega /c$ and
from Eq. (\ref{ratio}) it follows that at larger momenta the strength of the dual condensate
pole decreases quadratically in momentum relative to that of the second sound pole. Within
the confines of this gaussian treatment this makes again sense. At these short times and
distances one enters a regime where one is probing mainly the phase ordered matter forming
the background in which the vortices move.   This matter is the same stuff as the fully phase
ordered matter and accordingly it should carry the same Goldstone excitation. 
Eq. (\ref{mottprop}) tells how to interpolate between the disorder physics at $q \rightarrow 0$
with its number eigenstates and the phase ordered regime at large momenta: the dual
condensate pole looses its weight gradually, in fact in the same way as the Higgs mass
looses its influence on the dispersion.

\section {Dual view on the critical regime}

\begin{figure} 
\includegraphics[width=8.4cm]{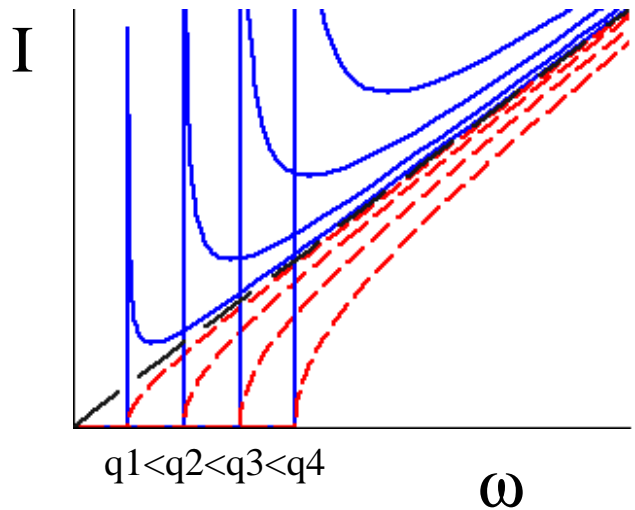}
\caption{The spectral functions associated with  the second sound (full lines) and condensate
(dashed lines) pieces of the velocity propagator in the critical regime  for various momenta
(Eq. \ref {ResponseCritical})). At $q = 0$ both critical continua  become degenerate and
linear in $\omega$ reflecting the simple correlation function exponent $\eta_A = 1$ associated with 
the dual gauge fields. However, at finite momenta it is seen that the second sound continuum diverges
at threshold while  the condensate piece is actually suppressed, although both continue to be 
governed by the same scaling dimension. At finite momenta this different behavior of the critical
continua has to be present in order for them to be consistent with the momentum dependence 
of the pole strengths of the propagating disorder excitations appearing at the moment one moves 
from the critical coupling.}  \label {FigCritical}
\end{figure}

\begin{figure} 
\includegraphics[width=8.4cm]{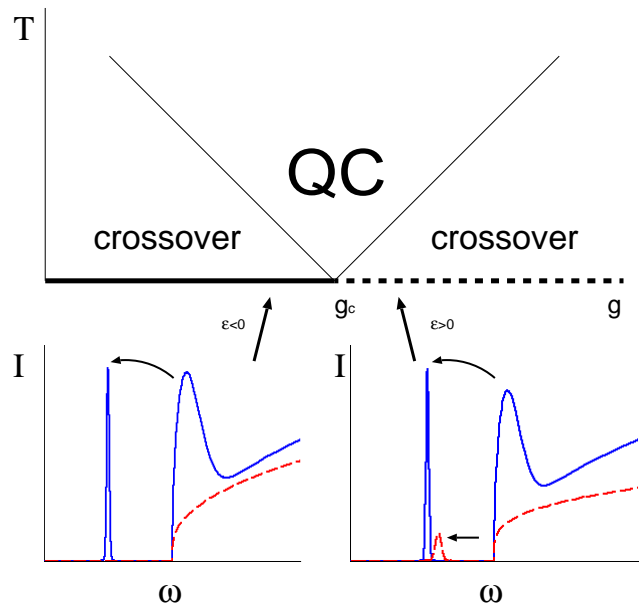}
\caption{Cartoon of the appearance of the second sound (full lines) and 
dual condensate (dashed lines) contributions to the velocity spectral functions
at finite momentum in the close vicinity of the quantum critical point, both on the ordered- (left)
and disordered  (right) side.  Although the critical continua are expected to have a 
very similar appearance, on the ordered side only a second sound pole is found. However,
on the disordered side the system scales to dual superconducting order with the effect that
one finds both a propagating second sound- and condensate excitations with strengths
governed by the XY correlation length exponent. However, the way
their pole strengths develops as function of momentum  tracks the gaussian result shown in
Fig. 2b and it turns out that the momentum dependence of the critical continua is just of the 
right kind to be consistent with the behavior of the disorder poles (section VI).} \label {FigCritClose} 
\end{figure}

We are not done yet. We have implicitly assumed up to this point that the fields are
non-interacting. Modulo perturbative corrections this would have been fine in dimensions
above the upper critical dimension but the Abelian-Higgs model in 2+1D is below its
upper critical dimension $d_{uc} = 3+1D$. One has now to be cautious with considerations
like the one in the previous section. Upon Exceeding the scale $\Omega$ one does not
re-enter the ordered phase but instead one enters the quantum critical regime which has
no longer to do with order or disorder but has acquired an own identity due to the strongly
interacting nature of the critical point. Away from the critical coupling, the order- (second sound)
and dual order (longitudinal and transversal photons) excitations discussed so far still make
sense because they will appear as bound states pulled out from the low energy side of the
continuum of critical modes, with a pole-strength and binding energy diminishing upon approaching
the critical coupling. The missing link at this point is the appearance
of the continua of critical modes as picked up by the velocity propagator.
After some preliminaries we will derive their form resting on the large
body of knowledge on the 3D XY critical state.  These critical continua
turn out
to behave in a quite surprising way, with the second sound- and condensate
contributions showing a completely different behavior away from $q = 0$
(see Fig. \ref {FigCritical} and Fig. \ref {FigCritClose}). We will subsequently focus in on the detailed
way the quasiparticle poles (second sound, the excitons) develop as
function of the distance from the critical point, making the case that
the critical continua have to be as they appear in order to be consistent
with the quasiparticle poles.

In order to describe the system close to- and right at  criticality, we introduce renormalized
parameters and  critical exponents. The role of  reduced temperature is taken by the
quantity $\epsilon = \frac {g - g_c}{g_c}$, which is the reduced coupling constant.
If $\epsilon <0$ or $\epsilon > 0$, we
approach the quantum critical point at $g_c$ from the ordered and disordered side, respectively. 
Pending if we approach  the critical point from the order- or disorder side,
the system will scale either to the stable fixed points associated with phase order and non-interacting 
second sound ($g=0$) or  with
non-interacting rotors ($g = \infty$). The reduced coupling constant $\epsilon$ is therefore
a relevant operator with scaling dimension $y_\epsilon>0$. Another relevant field,
which plays the role of the magnetic field in the standard scaling analysis is the
generating functional field ${\cal J}_\mu$. Since it is relevant at the transition, its
scaling dimension is also positive $y_{\cal J} > 0$.

The model we consider is relativistic, with dynamical critical exponent $z=1$, and  
its critical behaviour will coincide with
that of the 3D XY-model. We use the state of the art for the
exponents, based on analytic methods (high-temperature expansion
\cite {LeGuillouZinnJustin}, vortex-loop scaling\cite {Shenoy}, one-loop
renormalization group \cite {HerbutTesanovic}) as well as numerical results
from Monte Carlo simulations \cite {HoveSudbo, HasenbuschT, HasenbuschG}.
The critical exponent $\eta$ for the order parameter propagator
$\langle e^{i \phi_j} e^{-i \phi_i} \rangle$ has been studied in great detail
\cite {FisherWGF, Olsson, OlssonTeitel, HoveSudbo, HoveMoSudbo, FFHuse}.
However, our interest is in the velocity correlation function
Eq. (\ref {propagator_XY_nonrelativistic}) which  is not straightforwardly related to 
the vertex correlator away. Instead, we will use  the knowledge of the scaling 
dimensions of the dual field $\eta_{A}$ to derive the form of the velocity propagator
in the critical regime. 

Let us first analyse the model and its propagators right at the critical point $g = g_c$.
The exponent $\eta_A$ is usually defined as the critical exponent of the gauge
fields correlation function right at the critical point $g = g_c$, i.e.
$\langle \langle A A \rangle \rangle \propto 1/p^{2-\eta_A}$. 
To be consistent with the literature\cite{HoveSudbo, HerbutTesanovic},
we have to change the gauge fix from 'our'
Coulomb/unitary gauge fix to the Lorentz gauge fix ($\partial_\mu A_\mu = 0$,
i.e. the vector potential is purely transversal). In this gauge fix, the gauge field can
be projected onto a 3D linearly polarized basis (defined as ${\bf e}_0 = \frac {\bf p}p$,
${\bf e}_{-1} = - {\bf e}_T$ and ${\bf e}_{+1} = {\bf e}_{-1} \times {\bf e}_0$).
The component $A_0$ is set to zero by the gauge fix, with only the space-time
transversal components of the fields being
physical. The spatially transversal photon (second sound) degree of freedom $A_T$
is now represented by $A_{-1}$. The remaining component $A_{+1}$ that admixes
the Coulomb and the longitudinal photons plays the role of the vortex phase
degree of freedom in this particular gauge fix. On the Gaussian level of
the previous sections, the propagators for the gauge fields within the Lorentz gauge fix
degenerate and are given by
\be
  \langle \langle A_h^\dagger | A_{h'} \rangle \rangle = \rho_s \frac {\delta_{h, h'}}{\omega_n^2 + c^2 q^2 + \Omega^2} \label {propagator_A_Lorentz}.
\ee
In the Coulomb phase one finds  the same propagator with a vanishing gap, $\Omega = 0$.
The indices are taking `transversal' values
$h, h' = \pm 1$. The coupling constant in the prefactor is expressed in terms of the
superfluid stiffness $\rho_s = 1 / g$ which is a quantity which does renormalize. It follows
that the residues of the quasi-particle poles (order/disorder excitations) are also renormalized
which would not be the case if the prefactor would correspond with the bare coupling 
$g$. The overall prefactor in the expression for the  velocity
propagators corresponds with $g_b^2 \rho_s$ in this scheme. The $g_b^2$
is the bare critical coupling since the relation between the dual and original
propagators Eq. (\ref {Zaanen_Mukhin}) is an exact relation from the 
Legendre transformation, which is also valid  in the critical regime.
Accordingly, both the second sound of the ordered side and the excitons
of the disordered side loose their pole strength approaching the critical
point and this is governed by the renormalization of the superfluid density $\rho_s$
which we will deduce starting from the known critical behaviour of the dual gauge field
propagators.

Herbut and Te\v sanovi\' c \cite {HerbutTesanovic} analyzed the  charged XY model
which is  equivalent to the dual action Eq. (\ref {L_EM_full}). From their expression for
the $\beta$-function governing the renormalization of the electrical charge,
 it follows that at the fixed point
\be
  0 = \hat e_0^2 (D - 4 + \eta_A).
\ee
Assuming that the charge scales to a finite value $\hat e_0$, it follows that
$\eta_A = 4 - D \equiv 1$. The same result was obtained by Hove and
Sudb\o \cite {HoveSudbo}, using Monte-Carlo to  determine the
exponent $\eta_A$ from the vortex correlations at the critical point. They introduced
a relation between the correlation function of the vortex tangle $G(p)$ and the dual
gauge field propagator,
\be
  \langle {\bf A}^\dagger {\bf A} \rangle = \frac {2 \beta}{p^2} (1 - \frac {2 \beta \pi^2 G(p)}{p^2})
  \label {HSZaanenMukhin}
\ee
valid for the case of the uncharged original/charged dual action. Notice that in Ref. \cite {HoveSudbo}
${\bf h}$ is used for the dual gauge fields and ${\bf A}$ for the original gauge fields.
At the critical point the vortex correlator is given by
$\lim_{p \to 0} 2 \beta \pi^2 G(p) = p^2 - C_3 (g) p^{2+\eta_A} + \ldots$ and using
Eq. (\ref {HSZaanenMukhin}) it follows that
\be
  p^2 \langle {\bf A}^\dagger {\bf A} \rangle = C_3 (g) p^{\eta_A} + \ldots. \label {ppAA}
\ee
According to their numerical simulation, Eq. (\ref {ppAA}) shows a linear behaviour and 
Hove and Sudb\o \cite {HoveSudbo} conclude that $\eta_A = 1$.

The critical propagator of the dual gauge fields, Eq. (\ref {ppAA}), can  be used to establish 
the  form of the velocity propagator Eq. (\ref {propagator_XY_nonrelativistic}) in the critical
regime. Comparing Eq. (\ref {ppAA}) with our form of the gauge field propagator
Eq. (\ref {propagator_A_Lorentz}), and bearing in mind the degeneracy, we conclude 
that each field component propagator corresponds with one half of the propagator Eq. (\ref {ppAA}), 
\be
  \langle \langle A_h^\dagger A_{h'} \rangle \rangle = \frac {C_3}{2} p^{\eta_A - 2} \delta_{h, h'} + \ldots. \label {propagator_A_critical}
\ee
We can now use again the universal Zaanen-Mukhin relation Eq. (\ref {Zaanen_Mukhin})
to obtain the velocity propagator
\be
  \langle \langle v_i | v_j \rangle \rangle \sim P_{ij}^L \left \lbrack \frac {- \omega_n^2}{p^{2-\eta_A}} + \ldots \right \rbrack + P_{ij}^T \left \lbrack p^{\eta_A} + \ldots \right \rbrack \label {propagator_XY_critical}
\ee
right at $g = g_c$. This is the first main result of this section.
 The dots represent constant terms with no imaginary parts as well as
short distance corrections. At least deep in the critical regime the Wick rotation to real
time is simple\cite {SubirQPT} because scale invariance implies that Euclidean propagators are 
power laws, turning into branch cuts in real frequency 
With $\eta_A = 1$, right at the criticality, the spectral function has two quite different branch cuts in
the longitudinal and the transversal channel
\be
  {\rm Im} \langle \langle v_i | v_j \rangle \rangle_L &\sim& \theta (\omega^2 - c^2 q^2)
  \frac {\omega^2}{\sqrt {\omega^2 - c^2 q^2}}, \\
   {\rm Im} \langle \langle v_i | v_j \rangle \rangle_T &\sim& \theta (\omega^2 - c^2 q^2)
   \sqrt {\omega^2 - c^2 q^2},
  \label {ResponseCritical}
\ee 
and we sketch both pieces of the velocity correlator in Fig. (\ref {FigCritical}). $\theta (x)$ is
the Heaviside unit step function.

This is quite an unexpected result. At $q = 0$ we find both spectral functions
to be simply proportional to frequency, a simple behavior which of course
originates in $\eta_A = 1$. Upon increasing momentum, the 'sound' and 
'condensate' spectral functions start to behave very differently near
the threshold $\omega = c q$ although at large $\omega$ they merge together
again. The sound part shows the usual \cite {SubirQPT} divergence
$\omega^2 / ( \omega - c q )^{2 - \eta_A} \sim \omega^2 / ( \omega - c q)$ 
while the condensate piece develops like 
$( \omega^2 - c^2 q^2 )^{\eta_A/2}  = \sqrt{\omega^2 - c^2 q^2}$.
The degeneracy of the two contributions at $q = 0$ rings a bell: at
infinite wavelength it should be that the critical fluctuations are
eigenstates of rotor angular momentum, `equalizing' the condensate and
sound contributions as we found for the propagating excitations. To
understand better why these contributions should become different 
at finite momenta, we should first analyze in more detail what happens
with the quasiparticle poles close to the critical point.   

To analyze the behavior of the quasiparticle poles in the
ordered- and disordered phase close to the critical coupling. 
we need  hyperscaling. Although one 
has to be careful \cite {KBinder}, recent numerical
simulations \cite {HasenbuschG} show that there is none or a very small violation of the hyperscaling for 3D XY. 
Let us first repeat the standard hyperscaling arguments applied to the velocity-velocity propagators. 
We denote the propagators of the gauge field in real space as
$G_A (x, \epsilon)$. It is generated by the term ${\cal J}_h A_h$ 
in the action and this generating functional of the gauge fields ${\cal J}_h$ plays a role similar to a
magnetic field. It is a relevant field that scaling dimension $y_{\cal J}$.
Hyperscaling requires that such fields act on a block of $b^{d+1}$ points in space-time,
treated as a single variable. After a scale transformation,
the new propagator is related to the original one by
\be
  G_A \left ( \frac r b, {\cal J}' \right ) = \frac {\partial^2}{(\partial {\cal J}')^2} \ln Z \lbrack {\cal J} \rbrack
  \sim \frac {b^{2 (d+1)}}{\lambda_{{\cal J}}^2} G (r, {\cal J})
\ee
where $\lambda_{{\cal J}} = e^{y_{{\cal J}}}$ is the scaling factor of the generating functional for the gauge fields. 

Repeating the scale transformation $n$ times in the vicinity of the critical point, we obtain
\be
  G_A (r, \epsilon) = \frac {\lambda_{{\cal J}}^{2n}}{b^{2 n (d+1)}} G (r /b^n, \lambda_\epsilon^n \epsilon), \label {hyperscaling}
\ee
with the scaling factor $\lambda_\epsilon = e^{y_{\epsilon}}$ 
associated with the reduced coupling constant.
Choosing $n$ such that $(\lambda_\epsilon)^n b = const.$ from Eq. (\ref {hyperscaling})
it follows that the propagator behaves universally on both sides of the critical point as,
\be
  G_A (r,\epsilon)  \propto |\epsilon|^{\frac 2{y_\epsilon} (d+1-y_{\cal J})} \Phi_\pm (r / |\epsilon|^{-\frac 1{y_\epsilon}}). \label {G_hyper}
\ee
Te Functions $\Phi_\pm$ are universal functions associated 
with the ordered and disordered sides of the critical regime, 
and given in terms of $G (r, const)$. The denominator in its argument is the correlation length
that diverges at the critical point with exponent $\nu$, implying
the familiar relation
$\nu = 1 / y_\epsilon$. The relation of the `magnetic field' exponent $y_{{\cal J}}$ to the
scaling exponent $\eta_A$ follows when we set $\epsilon = 0$ in Eq.
(\ref {hyperscaling}),
\be
  y_{{\cal J}} = \frac {d+3-\eta_A}{2} \to 2
\ee
using the known value $\eta_A = 1$. Together with the relation
for $\nu$, Eq. (\ref {G_hyper}) can be written as
\be
  G_A (r, \epsilon ) &\propto& |\epsilon|^{\nu (d - 1 + \eta_A)} \Phi_\pm (r / |\epsilon|^{-\nu}) \\
  & \to& |\epsilon|^{2 \nu} \Phi_\pm (r / |\epsilon|^{-\nu}).
\ee
This is just the familiar result that the behaviour of the correlation function close to the critical point
is governed by the exponents $\nu$ and $\eta$ (with $\eta =1$
in the present case), and the crossover functions $\Phi_{\pm}$.

Let us first approach the critical point from the disordered side, i.e. $\epsilon \to 0^+$.
This phase is characterized by the gap Eq. (\ref {Omega}), which we can call
(compare Ref. \cite {SubirQPT}) $\Delta_+ = \Omega$. This gap is proportional to
the inverse correlation length of the vortex tangle $\xi = c / \Omega$.
Upon approaching the critical point, both the correlation length vanishes and
the gap diverges with characteristic exponent $\nu$ as $\xi \propto \epsilon^{-\nu}$
and $\Delta_+ \propto \epsilon^{z \nu}$, where $z = 1$ is the dynamical exponent which equals one in this specific case. 
The 3D XY correlation length exponent $\nu = 0.66-0.67 \approx \tfrac 2 3$ according to a large body of work
\cite {LeGuillouZinnJustin, Shenoy,HasenbuschT, HasenbuschG}. 
Given that there are two dynamical fields in the problem ($A_T$ and the vortex phase field phase $\phi_V$) one could be tempted 
to think that there are two correlation lengths in the problem,
but this is not the case the problem is effectively Lorentz
invariant, consistent with the numerical work\cite {NguyenSudbo, HoveMoSudbo, OlssonTeitel} and
an argument \cite {Olsson} linking it to the anomalous dimension of the gauge field $\eta_{A}$\cite {HerbutTesanovic, HoveSudbo}.

The scaling dimension of the superfluid density  can be 
deduced from
 Eq. (\ref {propagator_A_Lorentz}). After Fourier transformation
to space-time, the
Gaussian propagator Eq. (\ref {propagator_A_Lorentz}) behaves like
\be
  \langle \langle A_{h'}^\dagger | A_{h'} \rangle \rangle &=& \rho_s \frac 1{x^{d-1}} \Psi_+ (\frac x \xi) \\
  &=&
  \rho_s \xi^{-(d-1)} \Phi_+ (\frac x \xi) \\
  &=& \rho_s \epsilon^{\nu (d - 1)} \Phi_+ (\frac x \xi ).
\ee
Comparing it with the hyperscaling form for the gauge field propagator
Eq. (\ref {G_hyper}) we conclude that the superfluid density scales as
\be
\rho_s \propto |\epsilon|^{\nu (2 - \eta_A)} \to |\epsilon|^\nu
\ee
at the disordered side of the critical point. 

We have now 
arrived at a point where we can determine the behaviour of the 
two quasi-particle poles upon approaching
the critical point from the disordered side. Using Eq.
(\ref{propagator_XY_critical}), the fact that $g \rightarrow
g^2_b \rho_s$ and the scaling of both $\rho_s$ and $\Omega$,
we conclude  that the 
vortex-condensate pole $\sim P^T$ has a strength
proportional to $ \rho_s \Omega^2 \propto \epsilon^{2 z \nu + \nu (2 - \eta_A)}
\to \epsilon^{3 \nu}$, vanishing upon approaching the 
critical point with an exponent $3 \nu \cong 2$ while its strength
disappears in the critical continuum as indicated in
Fig. \ref {FigCritClose}c. Turning now to the second sound
pole $\sim P^L$, we observe that at long wavelength ($q 
\rightarrow 0$) its strength behaves exactly like the 
condensate pole. This has to be because eventually, at large
enough distances, one should recover the fact that these excitons
correspond with the exact rotor angular momentum eigenstates.
However, for increasing momenta the term in the numerator 
$\sim c^2 q^2$ takes over, and the strength of the
large momentum second sound pole is scaling more slowly to zero
upon  approaching the critical point, governed now by the
superfluid density exponent $\nu (2 - \eta_A) \cong \tfrac 2 3$.
This is of course not different from what we found on the
Gaussian level, with the second sound pole overtaking the
condensate pole when the vortex condensate is `loosing its grib',
governed by the Higgs mass $\Omega$.

To complete the picture, let us finally consider what happens
with the second sound pole approaching the critical point from
the ordered side. This is straightforward: as before, we should
substitute $g \rightarrow g^2_b \rho_s$ in the gaussian result 
Eq. (\ref {propagator_XY_nonrelativistic}) and $\rho_s \sim 
|\epsilon|^{\nu}$ because $\rho_s$ renormalizes in the same way 
on both sides of the transition\cite{josephsonlength}. In 
other words, the strength
of the second sound pole on the ordered side coincides with
its behavior at large momenta on the disordered side.  

Not surprisingly, we have found that the `order poles' behave
quite like the results we found on the Gaussian level in
the previous sections except that renormalized mass scales and -quasiparticle
residues have to be used, all governed by the same correlation exponent $\nu$ ,
because $\eta_A$ `magically' drops out. We can now use this
knowledge to comprehend why the critical continua of
Fig. \ref {FigCritical} behave the way they do. We already
argued that at energies far away from the threshold $\omega
= c q$ the second sound and vortex condensate pieces picked
up by the velocity correlator merge in the same linear  $I \sim 
\omega$ behavior. At finite $q$ the differences between the two are large near
threshold. With the knowledge regarding the behavior of the
quasiparticle poles at hand this now makes sense. $\rho_s$
being a relevant operator, its influence at high energies is
small while growing when times get longer. A bit away from
the critical point, it takes over at a length $\sim \xi$ where
the system gets under control of the stable fixed points at
zero or infinite coupling, which are also in charge of 
protecting the  quasiparticle poles.  Surely, the
quasiparticles close the critical point can be viewed as
bound states pulled out of the critical continuum due to
the effect of the relevant operators (Fig. \ref {FigCritClose}). However, because of
the way the latter scale, the quasiparticles are formed from
the {\em low energy end} of the critical continuum. What does 
this mean for our velocity propagator, 'watching' the true 
critical excitations through the `duality filter'? We derived
some clear rules for how the weights should be distributed
over the quasiparticles: the condensate and sound poles of
the disordered state should have equal weight at $q=0$, but 
the former should loose its weight rapidly for increasing 
momentum. Inspecting now the low energy end of the critical
continua for various momenta we see this rule also at work
(Fig. \ref {FigCritical})! We notice that this 'weight-matching'
of the critical continua and the quasiparticle poles is
to an extent even quantitative. For this purpose we inspect
the pole strength ratio Eq. (\ref {ratio}) close to the critical point.
For fixed $q$, due to the gap in the denominator, we learned already that the ratio diverges
like $\sim q^2/\epsilon^{2 \nu}$. However, the prefactor
of the second sound pole strength is however proportional to $q^2$. Comparing it with
the ratio of the spectral responses right at the critical point and near threshold
($\omega \approx c q$)
\be
  \left ( \frac {I_L}{I_T} \right)_{g_c} = \frac {\omega^2}{\omega^2 - c^2 q^2} \stackrel {\omega \approx c q}{\longrightarrow} q^2 \times \mbox {``divergent part''}.
\ee
We find a perfect match -- the strengths of the spin-wave and the condensate
excitations are proportional to the strengths of their respective critical continua where they
have their 'origin'.

Surely, this does not explain everything, and to a degree
Eq. (\ref{ResponseCritical}) is a result which stands on its own. However,
given the simple integer $\eta_A$ exponent, it appears 
to us to be a unique analytical form which obeys 
the general requirements of scale invariance
and Wick rotation, having at the same time the right form
to be consistent with the evolution of the spectral weights
in the quasiparticle poles.

\section {Conclusion}

What have we learned? We have taken the simple example of phase
dynamics at zero chemical potential to predict the form of the
superfluid velocity correlator in the ordered, disordered and
critical regimes, exploiting the vortex duality. Although it
does not seem to be generally recognized, the dual order of
the disordered side does manifest itself when the system
is interrogated with 'orderly means'. In our example, one
of the two degenerate excitons of the phase disordered/Mott 
insulating state can as well be called the longitudinal photon
associated with the phase-rigidity of the dual superconductor. Although
less obvious in the present simple example, this might be used as 
a technical convenience. Quite generally, it is easier to compute
the excitation spectrum of the system 'around' the ordered state,
helped by the Goldstone theorem, the Higgs mechanism, etcetera,
while disorder is not a convenient starting point when the interactions
are strong. It is actually so that this work was originally inspired
on  problems encountered in the study of quantum liquid crystals
where this `dual route' to the spectral functions associated with
measurable quantities seems to be the only way available\cite{ZMN}.

This dual route also tells another story which is far less obvious.
This can be summarized as: `studying the disordered state with order
operators, one recovers the signal of the ordered state at energies
and momenta where the dual order parameter looses control.' In our
specific example, the dual condensate piece of the exciton doublet
of the disordered state fades away when momentum is increased and
at large momentum only second sound remains. We argued that this
same 'mechanism' is even at work in the critical state, being
ultimately responsible for the rather odd appearance of the continuum
of critical modes as measured by the velocity correlator.

Within the field-theory this is surely correct -- is is based on
controlled calculations. Another issue is, can we literally apply
the field theory to condensed matter problems for this particular
purpose? We are actually not sure. One way to read the effect of
the previous paragraph is as follows. It is assumed in the field 
theory that the lattice constant is vanishing. As applied to any
problem with a finite lattice cut-off, this means that the 
field theory can only be taken literally when the distance between
vortices is large compared to the lattice constant -- the small
fugacity limit is implicitly taken. If this is the case, by zooming in
one will eventually get at length scales which are smaller than
the inter-vortex distance, and here one will rediscover the
implacable order and  its dynamical implication  in the form of its Goldstone 
mode. The way this limit is reached is a bit more sophisticated than
suggested with these words, but this we discussed at length in this
paper.

The problem is  that in real condensed matter systems the vortex
(or, in general, 'dual') matter is actually rather dense when density is measured 
in units of the lattice constant. Although the field theory is fine
in the scaling limit, as helped by universality, to what extent can
the effects we discussed in this paper become noticeable? We leave this
as an open question, in the hope that others might get motivated to
have a closer look.

\section {Acknowledgments}

We acknowledge helpful discussions with S.I. Mukhin and
D. Nogradi. This work was supported 
by the Netherlands foundation for fundamental research 
of Matter (FOM).

\bibliographystyle{apsrev}

\begin{thebibliography}{10}
\bibitem {josephsonlength}
Notice that the Josephson correlation length scales according
to,  
\be
  \xi_J = c / \rho_s \propto |\epsilon)|^{- \nu (2 - \eta_A)} \stackrel {\eta_A = 1}{\longrightarrow}
  |\epsilon|^{-\nu}. \label {xiJ_scaling}
\ee
It is well known\cite {FisherBarberJasnow} that $\xi_J$ is 
governed by the same exponent $\nu$ in
charge of the mass on the disordered side. According to the
above equation, this can only be the case when $\eta_A =1$,
once again highlighting the remarkable consistency of
the duality construction!
\end{thebibliography}

\end {document}